\documentclass{iau} 
\usepackage{graphicx}

\title[The Early Merger that Made the Stellar Halo] 
{The Early Merger that Made \\ the Galaxy's Stellar Halo}

\author[N. Wyn Evans]   
{N. Wyn Evans$^1$}

\affiliation{$^1$Institute of Astronomy, Madingley Rd, Cambridge CB3 0HA\\}

\pubyear{2019}
\volume{353}  
\jname{Galactic Dynamics in the Era of Large Surveys}
\editors{M. Valluri, \& J. A. Sellwood  }
\begin{document}

\maketitle

\begin{abstract}
The last two years have seen widespread acceptance of the idea that the Milky Way halo was largely created in an early (8-10 Gyr ago) and massive ($> 10^{10} M_\odot$) merger. The roots of this idea pre-date the Gaia mission, but the exquisite proper motions available from Gaia have made the hypothesis irresistible. We trace the history of this idea, reviewing the series of papers that led to our current understanding.
\end{abstract}

\firstsection 
\section{Before Gaia}

In a premonitory paragraph, \cite[Gilmore \& Wyse (1985)]{Gi85}  wrote {\it \lq\lq a possible model for the Galactic thick disk is that the Galaxy suffered a collision with a companion galaxy shortly after the onset of formation of the Galactic (thin) disk, or more simply that Galactic formation involved the infall of sizable fragments over a time scale longer than that required to make the thin disk a reasonable entity. Simple calculations imposing conservation of angular momentum, linear momentum, and energy indicate that capture of a galaxy or fragment of mass $\sim 10^{10} M_\odot$ , about equal to the mass of the LMC, from an orbit with an apogalacticon of around 50 kpc would result in an input of energy sufficient to puff up any existing thin disk."} This is very close to what we now believe happened to the early Milky Way.

The germ of this idea can be traced to influential simulations of sinking satellites by \cite[Quinn \& Goodman (1986)]{Qu86}. They showed that thin disks become thicker and hotter as infalling satellites couple to the radial and vertical motions of thin disk stars. If disks are to maintain their remarkable thinness over a Hubble time in the presence of repeated satellite infall and merger, it may be necessary to reform the thin disk over the history of a galaxy. They suggested one possible source of new material to form such a new disk may be the gas provided by the satellite. This idea was subsequently confirmed and extended by many later numerical studies \cite[(e.g. Walker et al 1996, Velasquez \& White 1999, Villalobos \& Helmi 2008)]{Wa96, Ve99, Vi08}.

From an observational perspective, the dichotomy in the properties of the Milky Way's stellar halo is a long-standing theme~\cite[(e.g., Chiba \& Beers 2000, Carollo et al 2007, Nissen \& Schuster 2010)]{Ch00,Ca07,Ni10}. These papers claim that there are two distinct populations in the stellar halo; an inner, metal-rich and mildly prograde one and an outer, metal-poor and retrograde one. The evidence is mainly based on spectroscopic data -- radial velocities, metallicities and abundances. Without the benefit of accurate distances and proper motions from Gaia, the statements on kinematics in these papers are preliminary. However, we now know that the dichotomy in the stellar halo is truly present. The metal-rich inner halo can now be recognized as nothing other than the residue of the early merger event. 

Within the $\Lambda$CDM Cosmology, the stellar halo of a Milky Way-like galaxy is predicted to be dominated by a small number of massive accreted dwarf galaxies \cite[(e.g. Bullock \& Johnston 2005, Font et al 2006)] {Bu05, Fo06}. \cite[Deason et al (2013)]{De13} put forward the hypothesis that
the rapid transition in the Galactic stellar halo structural properties at break radius of 20-30 kpc
\cite[(e.g. Watkins et al 2009, Deason et al 2011)] {Wa09,De11} is likely associated with the apocentric pile-up of a relatively early (8-10 Gyr ago) and single massive accretion event. This prediction has now been vindicated.

All these papers contain ingredients of the truth, but they failed to fully convince the community. However, once data from \cite[the Gaia Collaboration (2016a)]{Ga16} was released, the evidence for an early merger became overwhelming. We will trace the development of this idea through the Gaia data releases.

\begin{figure}[t]
\begin{center}
 \includegraphics[width=5.2in]{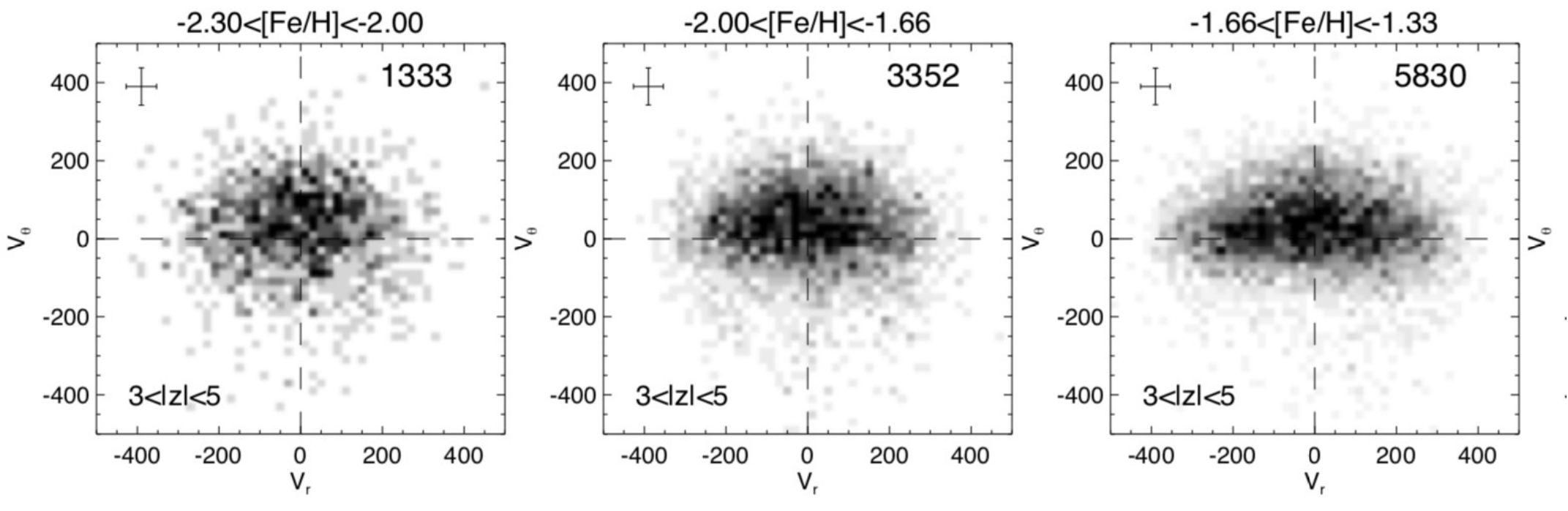} 
 \caption{The \lq Gaia Sausage\rq. This shows typical behaviour of stellar halo stars, here at distances $-2.3 < |z| < 5$ from the Galactic plane. Note that the velocity distribution ($v_r, v_\theta$) transforms from round to sausage-shaped as we move from left to right, or from lower metallicity to higher metallicity (From Belokurov et al 2018).}
   \label{fig:EvansFigOne}
\end{center}
\end{figure}

\section{Gaia Data Release 1}

\cite[The Gaia Collaboration (2016b)]{Ga16b} made Data Release 1 (DR1) available on 14 September 2016. The source catalogue contained positions of $\approx$ 1.1 billion sources. The astrometric catalogue was restricted to $\approx$ 2 million of the brightest stars in common with the Hipparcos and Tycho-2 catalogues, though very few of these stars belonged to the halo.

\cite[Belokurov et al (2018)]{Be18} hit upon the idea of cross-matching the Gaia DR1 source catalogue with the earlier Sloan Digital Sky Survey Data to derive proper motions. This gave the SDSS-Gaia catalogue of $\sim 80,000$ main sequence turn-off halo stars in a seven-dimensional phase space of three positional coordinates, three velocity coordinates and metallicity. With these proper motions in hand, \cite[Belokurov et al (2018)]{Be18} showed the striking variation in orbital properties of halo stars on moving from low to high metallicity, as illustrated in Fig.~\ref{fig:EvansFigOne}. At the low [Fe/H] end, the orbital anisotropy is almost isotropic. However, for halo stars with [Fe/H]$> −1.7$, the orbits become very radial and so the velocity distribution becomes extented or sausage-like (the so-called \lq Gaia Sausage\rq). Using cosmological zoom-in simulations, \cite[Belokurov et al (2018)]{Be18} argued that the extreme radial anisotropy is inconsistent with the continuous accretion of many dwarf satellites. Instead, the stellar debris in the inner halo was deposited in a single, head-on, accretion event by a satellite with $M > 10^{10} M_\odot$ between 8 and 11 Gyr ago. The radial anisotropy is caused by the dramatic radialisation of the massive progenitor\rq s orbit, as first seen in simulations by \cite[Amorisco (2017)]{Am17}.
  
\cite[Myeong et al (2018a)]{MyAS} used the same SDSS-Gaia cross-match to examine the Milky Way halo in action space ($J_R, J_\phi, J_z$). They found that the metal-rich population or the \lq Gaia Sausage\rq\ is more distended toward high radial action $J_R$ as compared to azimuthal or vertical action, $J_\phi$ or $J_z$. It has a mild prograde rotation, is radially anisotropic and highly flattened, with axis ratio $q \approx 0.6-0.7$. The metal-poor population is more evenly distributed in all three actions with a mild radial anisotropy, and a roundish morphology ($q \approx 0.9$). 

So, Gaia DR1 already established the narrative that the metal-rich halo stars may have come from a single accretion event. Let us see how Gaia DR2 embellished the story.

\begin{figure}[t]
\begin{center}
 \includegraphics[width=2.5in]{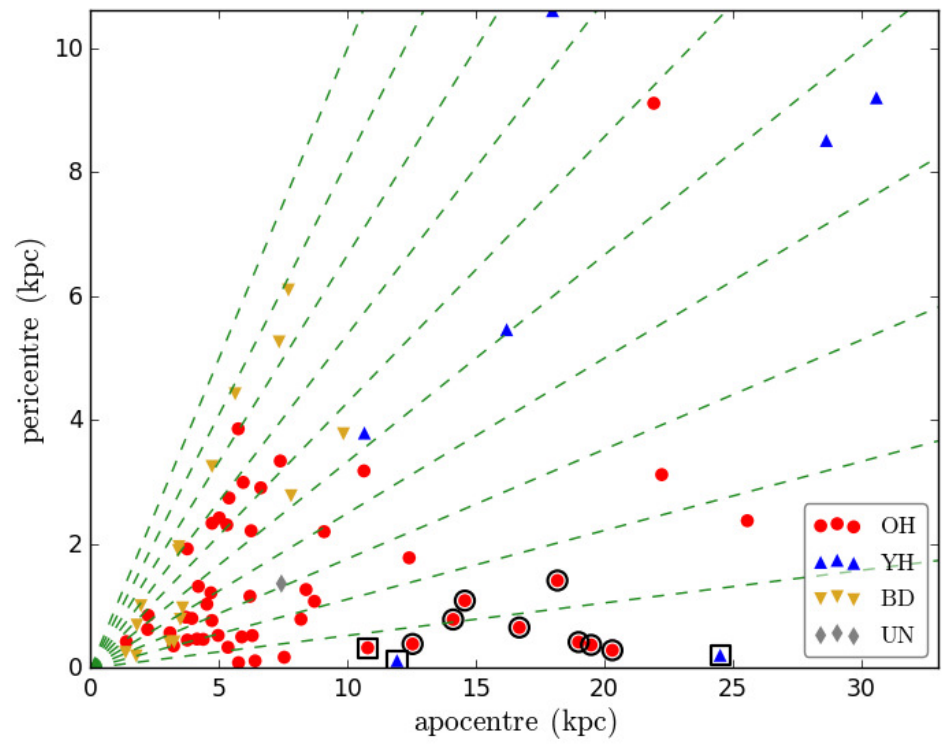}
 \includegraphics[width=2.5in]{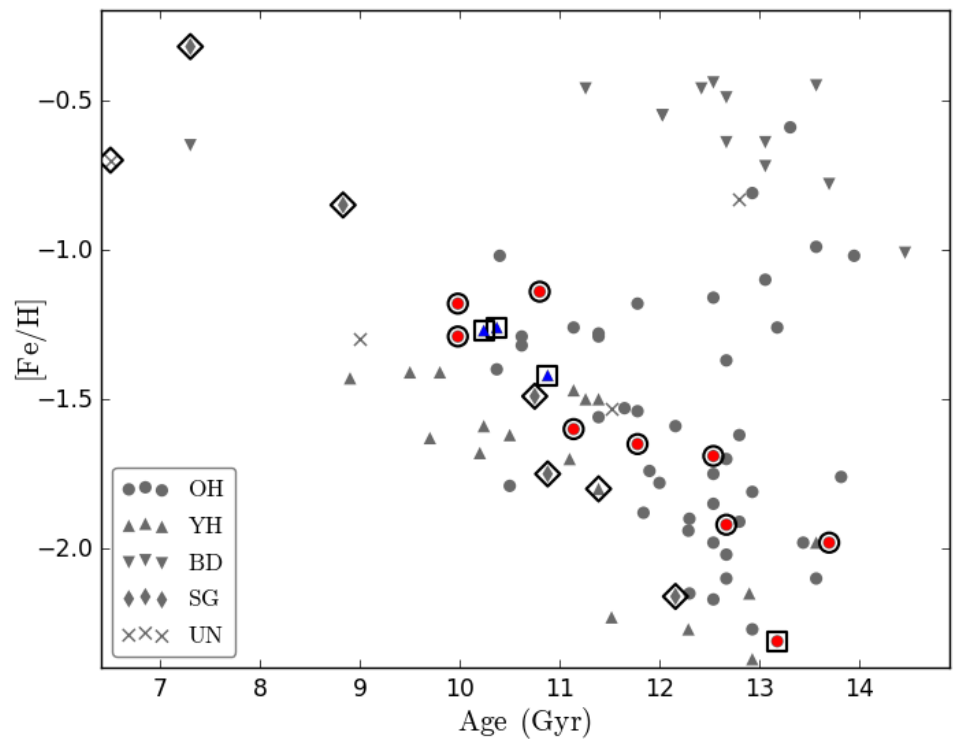} 
 \caption{The Sausage Globular Clusters. Left: The old halo (OH), young halo (YH), bulge/disk (BD) and unclassified (UN) globular clusters (GCs) plotted in the plane of pericentre versus apocentre.  Lines of constant eccentricity from 0 to 0.9 in steps of 0.1 are shown as dashed. Note the Sausage GCs (black open circles as probables and open squares as possibles) all have eccentricity $\gtrsim 0.80$. This is a population of highly radially anisotropic GCs that came in with the Sausage galaxy. Right  Plot of the age of GCs versus metallicity. The Sausage GCs are shown with circular (probables) and square (possibles) black boundaries. 7 GCs that are former denizens of the Sagittarius (Sgr) dwarf are shown as unfilled black diamonds. The sequences of Sgr GCs and Sausage GCs lie on two distinct, though closely matched, tracks. They are different from the bulk of the Milky Way GCs which show a constant age of $\sim 13$ Gyr independent of metallicity. (From Myeong et al 2018b)}
   \label{fig:EvansFigTwo}
\end{center}
\end{figure}

\section{Gaia Data Release 2}

\cite[The Gaia Collaboration (2018a)]{Ga18} made Data Release 2 (DR2) available on 25 April 2018. This was a transformational dataset that increased the number of stars with parallaxes and proper motions by at least three orders of magnitude. DR2 provided positions and the apparent magnitudes for $\approx$ 1.7 billion sources. For $\approx$ 1.3 billion of those sources, parallaxes and proper motions are available. However, only for the $\approx 7$ million stars brighter than $G \approx 12$ are radial velocities available from Gaia's Radial Velocity Spectrometer. Cross-matches with ground-based spectroscopic surveys (like APOGEE or LAMOST) remedy this deficiency.

\cite[Myeong et al (2018b)]{My18GC} quickly realized that any massive satellite must itself have contained its own population of globular clusters. They looked for the associated Sausage Globular Clusters (GCs) by analyzing the structure of 91 Milky Way GCs in action space using the Gaia DR2 catalog, complemented with Hubble Space Telescope proper motions. They identified 8 high-energy, old halo GCs strongly clumped in azimuthal and vertical action, yet strung out at extreme radial action. As shown in the left panel of Fig.~\ref{fig:EvansFigTwo}, they are very radially anisotropic and move on orbits that are highly eccentric ($e \approx 0.80$). They also form a track in the age-metallicity plane compatible with a dwarf galaxy origin. As the right panel of Fig.~\ref{fig:EvansFigTwo} shows, it is similar to, but slightly offset from, the track of the Sagittarius GCs. These properties are all consistent with the picture that the merger event that gave rise to the Gaia Sausage brought in an entourage of GCs. In fact, without using Gaia data at all, \cite[Kruijssen et al (2019)]{Kr19} identified the same group of GCs (which they attributed to an event called Canis Major). A larger catalogue of GCs with proper motions was subsequently constructed by \cite[Vasiliev (2019)]{Va19}. The membership of the Sausage GCs was then extended to $\sim 20$ by \cite[Myeong et al (2019)]{My19} using this enhanced catalogue.

As part of DR2, \cite[the Gaia Collaboration (2018b)]{Ga18} presented an intriguing Hertszprung-Russell (HR) diagram for stars with tangential velocity $v_{\rm T}  > 200$ kms$^{-1}$, which possessed two main sequences and two turn-offs. \cite[Haywood et al (2018)]{Ha18} looked at the chemistry of the blue and the red main sequences to understand this dichotomy. To do this, they cross-matched Gaia DR2 with the APOGEE stars \cite[(Abolfathi et al 2018)]{Ab18}, which have radial velocities, metallicities and abundances. They found that the red sequence is dominated by thick disk stars with metallicities between $−0.4>$ [Fe/H] $> −1$. The blue sequence stars typically have large apocenters. They are chemically consistent with a low star formation efficiency sequence, typical of a massive dwarf galaxy with metallicity extending from [Fe/H] $\approx -0.7$ to  -2.0. They concluded: {\it \lq\lq Belokurov et al (2018) suggested that the majority of the halo stars within 30 kpc are the remnant of a massive satellite accreted during the formation of the Galactic disk between about 8 and 11 Gyr ago. Our analysis supports this scenario: (1) the blue sequence stars seem to constitute a significant fraction of all stars with high transverse velocities; (2) we show that they could be on a chemical evolutionary track that is less $\alpha$-enriched than disk stars at the same metallicity and thus are compatible with an accreted population.\rq\rq} The first analysis of the chemical evidence complemented the earlier dynamical evidence to present a compelling picture of an early merger event.

\begin{figure}[t]
\begin{center}
 \includegraphics[width=5.5in]{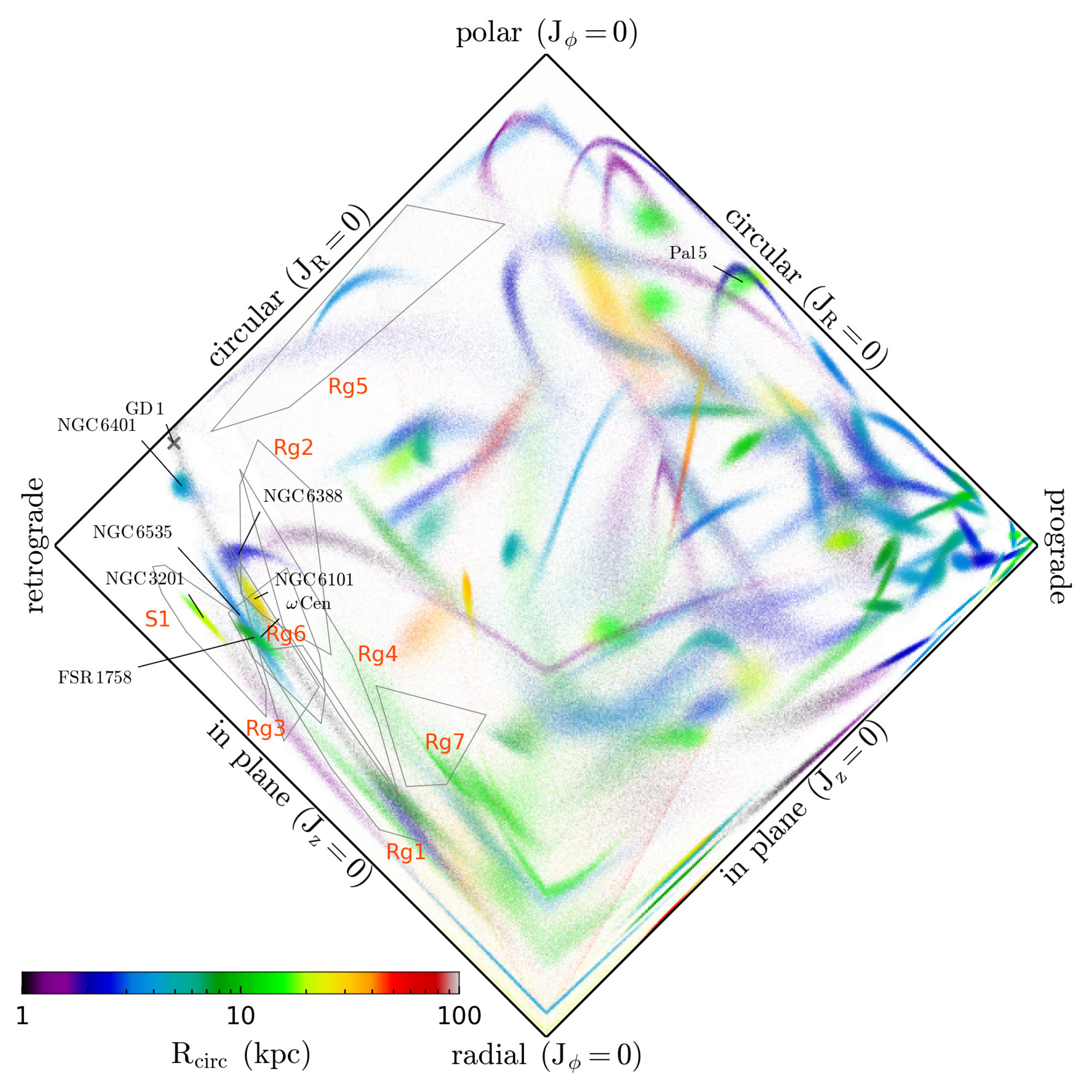}
 \caption{The action-space map for the Milky Way GCs and retrograde substructures. The horizontal axis is $J_{\phi} / J_{\mathrm{tot}}$, and the vertical axis is $(J_z - J_R) / J_{\mathrm{tot}}$. Each object is shown with 1000 Monte Carlo representations of the orbit as drawn from the observational errors. The shading marks the circular orbit radius for the corresponding total energy $R_{\mathrm{circ}}(E_\mathrm{tot})$. Objects clustered in this plot with similar shading are candidates for accretion events (From Myeong et al 2019).}
   \label{fig:EvansFigThree}
\end{center}
\end{figure}

Subsequently, \cite[Helmi et al (2018)]{He18} also studied the two distinct sequences in the Gaia HR diagram. They laid emphasis on the retrograde nature of kinematic substructures they had earlier found near the Sun~\cite[(Koppelman, Helmi \& Veljanoski 2018)]{Ko18}, arguing that they are of accreted origin. Again using APOGEE data, they analysed the kinematics, chemistry and spatial distribution of stars in the thick disk and the stellar halo. They concluded that the inner halo is dominated by debris from a single object that at infall was more massive than the Small Magellanic Cloud, which they nicknamed \lq\lq Gaia-Enceladus". They estimated the merger event had a mass ratio of four to one, and created the Galactic thick disk by dynamical heating $\approx 10$ billion years ago. Gaia-Enceladus has some similarities with the Gaia Sausage. Both theories envisage a major accretion event by a satellite with mass $> 10^{10} M_\odot$ between 8 and 10 Gyr ago. However, there are substantial differences between the details of the theories that are amenable to observational tests: (1) the Gaia Sausage is the result of a head-on collision, whereas the Gaia-Enceladus is not a head-on collision, but produces prograde and retrograde stars over a swathe of angular momentum with $−1500 < J_\phi < 150$ kpc kms$^{-1}$ (see Figure 2 of \cite[Helmi et al 2018]{He18}); (2) the stars in the Gaia Sausage have zero or possibly mild net prograde rotation, whereas the Gaia-Enceladus stars have strong net retrograde rotation; (3) the spatial structure of Gaia-Enceladus shown in Figure 3 of \cite[Helmi et al 2018]{He18}) is different from the Gaia Sausage, whose triaxial form has been traced in RR Lyrae stars \cite[(Iorio \& Belokurov 2019)]{Io19}; (4) there are no chemical differences between the bulk of the retrograde and zero angular momentum high energy stars in the Gaia-Enceladus theory, but there are for the Gaia Sausage.

\cite[Myeong et al (2019)]{My19} argued on dynamical grounds that the retrograde stars identified as members of Gaia-Enceladus by \cite[Helmi et al (2018)]{He18} are likely to be the residue of a different, but prominent, accretion event. This was already hinted at earlier by \cite[Bekki \& Fremann (2003) and Majewski et al (2012)]{Be03Ma12}. As actions are conserved under evolution, stars accreted in the same merger event are clustered in action space. \cite[Myeong et al (2018a,c)]{MMyAS,MyR} devised an algorithm to search for stellar over-densities in action space with respect to the data-derived background model, validating it against mock catalogues of substructures from the Aquarius simulations. They then identified a series of high significance retrograde substructures in the SDSS-Gaia data, labelled Rg1 to Rg7. Fig~\ref{fig:EvansFigThree} shows a projection of action space, from which we notice that many of the retrograde substructures (Rg1-Rg4 and Rg6) are coincident with 6 retrograde globular clusters. This includes two enormous ones, FSR 1758 and $\omega$ Centauri — long thought to be the core of a disrupted galaxy. This is consistent with the wreckage of another dwarf galaxy, less massive than the Sausage, which came in on a strongly retrograde orbit, disgorging globular clusters and debris through the inner Galaxy. \cite[Myeong et al (2019)]{My19} named this the \lq Sequoia Event'.

\begin{figure}[t]
\begin{center}
 \includegraphics[width=4.5in]{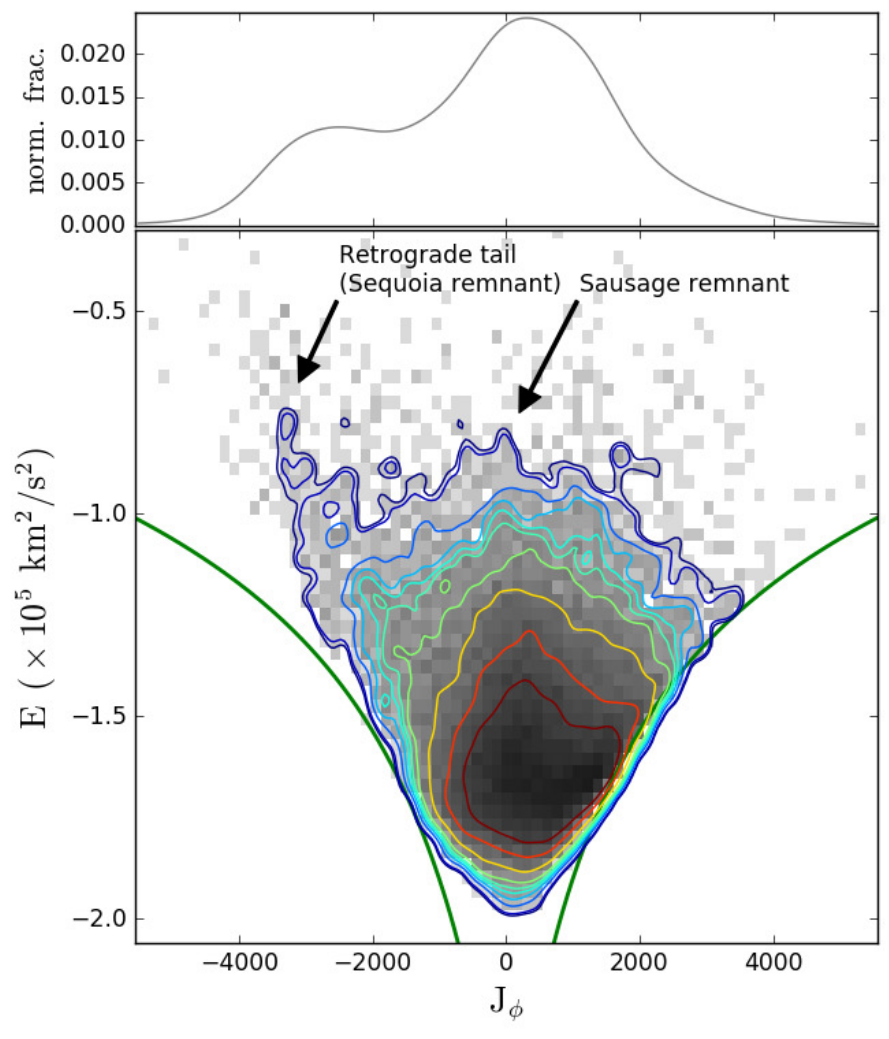}
 \caption{Distribution of energy and azimuthal action for the halo stellar sample similar to Figure~2 of \cite[Myeong et al (2018a)]{MyAS}. The top panel shows the distribution function of the azimuthal action for the stars with high energy ($E>-1.1\times10^5$~km$^2$s$^{-2}$). The signature of Gaia Sausage remnants is visible as a peak at low $J_{\phi}$. A separate trace of a retrograde accretion (Sequoia event) is clearly visible as a sharp tail around $J_{\phi}\sim-3000$~kpc~kms$^{-1}$. The boundary lines mark the circular orbit. (From Myeong et al 2019)}
 \label{fig:EvansFigFour}
\end{center}
\end{figure}

Of course, $\omega$ Centauri has long been one of the \lq usual suspects\rq\ \cite[(Bekki \& Freeman 2003)]{Be03}. It has a present-day mass of $5 \times 10^6 M_\odot$. It has multiple stellar populations. Its stars exhibit a large metallicity spread and there are extreme star-to-star variations in many light elements. If it was a dwarf galaxy, then its mass may once have been $\sim 10^{10} M_\odot$, based on models of chemical evolution of multi-population clusters. FSR 1758 was studied by \cite[Barba et al. (2019)]{Ba19} using DECaPS and VVV data, complemented with Gaia DR2.  It is an unusually extended GC, located at ($\ell = 349^\circ, b = 3^\circ$) and with a distance of ~10  kpc. It is also very retrograde. Whether one of the enormous GCs is indeed the core of the Sequoia, or whether they are merely large GCs once contained within it, remains an open question.

The total mass of the Sequoia galaxy was at least $1\times10^{10} M_\odot$, judged from the current mass of the associated GCs. \cite[Valcarce \& Catelan (2011)]{Va11} suggested the mass of the progenitor of $\omega$~Centauri may be as high as $10^{10} M_\odot$ from chemical evolution modelling of its multiple populations, which is in broad agreement. The mass-metallicity relation of \cite[Kirby (2013)]{Ki13} with a metallicity of $-1.6$ gives a broadly consistent stellar mass of $2\times10^{7} M_\odot$. Though less massive then the Gaia Sausage, the Sequoia was a notable accretion in the evolutionary history of the Milky Way. In terms of the stellar and total mass, the Fornax dwarf spheroidal could be a rough representation of the Sequoia progenitor. The fact that the Fornax dSph hosts a comparable number of GCs adds to the similarity. 

Are there chemical differences betwenn the Gaia Sausage and the Sequioa stars? With APOGEE DR14, \cite[Mackereth et al (2019)]{Ma19} found that halo stars with high eccentricity orbits tend to have lower [Mg/Fe] on average compared to the rest of the halo stars. More specifically, \cite[Myeong et al (2019)]{My19} showed that stars in the highly radial Gaia Sausage have different metallicity distributions and abundance patterns from stars in the high energy retrograde substructures. The Sausage stars show a metallicity distribution function peak at [Fe/H]$=-1.3$, whereas the high energy retrograde stars are more metal-poor, with a peak at [Fe/H]$=-1.6$. While the metallicitiy distributions of the Sausage and Sequoia stars overlap,  the two galaxies show distinct patters in the abundance of alpha elements. For example, at [Fe/H]$\sim-1.5$, the Sequoia debris are more enhanced in Al compared to the Sausage. \cite[Matsuno et al (2019)]{Mat19} searched through a database of $\sim 880$ metal-poor stars with [Fe/H] $< -0.7$. They also found that the high energy retrograde stars are distinct from the zero-angular momentum stars of the Sausage, which dominates the inner halo.  They reported that the `knee' in the abundance and metallicity plane differs by about 0.5 dex (at [Fe/H]$\sim-2$ for the Sausage and $\sim-2.5$ for the retrograde stars), which is another indication of their different origin. 

The Sequoia Event is discernible both dynamically and chemically. It forms a separate grouping from the bulk of the Gaia Sausage, which has close to zero net angular momentum. This is illustrated in Fig.~\ref{fig:EvansFigFour}, where the morphology of the contours in the high energy region shows a pattern of bimodal accretion tracks. The upper panel of Fig.~\ref{fig:EvansFigFour} shows the distribution of azimuthal action for the stars with high energy (e.g., $E>-1.1\times10^5$~km$^2$/s$^2$). The existence of this extra retrograde component, clearly separated from the Sausage at zero angular momentum, is evident. The signal is concentrated at a specific range of metallicity ([Fe/H] $\sim-1.6$). Thus, the Sequoia Event is also distinct from the Gaia-Enceladus structure, which appears to combine parts of the Gaia Sausage and the Sequoia.

\section{Conclusions}

The roots of the early merger event reach a long way back~\cite[(e.g., Gilmore \& Wyse 1995, Chiba \& Beers 2000, Carollo et al 2007, Deason et al. 2013)]{}. The real \lq star of the show' though is data from the Gaia satellite, which has been indispensable in convincing us that much of the Milky Way's stellar halo was built in a single massive ($>10^{10} M_\odot$) early (8-10 Gyr ago) encounter that probably also created the thick disk and replenished the gas to make thin disk stars anew.

The Gaia Sausage is the biggest structure in the stellar halo, with a stellar mass of $\sim 5-50 \times 10^8 M_\odot$. This is comparable to the stellar mass of the Large Magellanic Cloud. The total mass is probably $\sim 10^{11} M_\odot$. It is composed of strongly radial stars, indicative of a head-on encounter. A smaller merger, the Sequoia event, produced the strongly retrograde material in the stellar halo. It has a stellar mass of $\sim 2  \times 10^7 M_\odot$. This is the stellar mass of Fornax. The total mass is  $\sim 10^{10} M_\odot$.

The Gaia-Enceladus is not the same as the Gaia Sausage. It is not produced in a head-on merger. Its stars range in angular momentum from prograde through zero to extreme retrograde. There are observational tests that can be made to distinguish between the two theories. Specifically, high resolution spectroscopy will reveal whether the chemistry of the extreme retrograde halo stars and the eccentric stars is consistent with a single dwarf galaxy origin.

\end{document}